\documentclass[aps,prc,showpacs,preprintnumbers]{revtex4}%
\usepackage{amssymb}
\usepackage{amsmath}
\usepackage{amsfonts}%
\usepackage{graphicx}
\usepackage{dcolumn}
\usepackage{bm}

\begin{document}

\title{Tunneling of a composite particle: Effects of intrinsic structure}
\author{C.A. Bertulani$^{1}$, V.V. Flambaum$^{2}$,
and V.G. Zelevinsky$^{3}$}
\address{ $^1$ Department of Physics, Texas A\&M University-Commerce, Commerce, TX
75429,
USA\\
$^{2}$School of Physics, University of New South Wales, Sydney
2052, Australia \\
$^{3}$National Superconducting Cyclotron Laboratory and\\
Department of Physics and Astronomy, Michigan State University, East
Lansing, MI 48824-1321, USA}

\begin{abstract}

We consider simple models of tunneling of an object with intrinsic
degrees of freedom. This important problem was not extensively
studied until now, in spite of numerous applications in various
areas of physics and astrophysics. We show possibilities of
enhancement for the probability of tunneling due to the presence of
intrinsic degrees of freedom split by weak external fields or by
polarizability of the slow composite object.
\end{abstract}

\date{\today}

\maketitle

Quantum tunneling is a subject of constantly renewed interest, both
experimentally and theoretically. The standard textbook approach
describes the tunneling process for a point-like particle in an
external static potential. Chemical and nuclear subbarrier reactions
\cite{balantekin98}, especially in astrophysical conditions, as a
rule, involve complex objects with their intrinsic degrees of
freedom. As stated in Ref. \cite{saito94}, ``Although a number of
theoretical works have studied tunneling phenomena in various
situations, quantum tunneling of a {\sl composite} particle, in
which the particle itself has an internal structure, has yet to be
clarified." There are experimental data \cite{arzhan91,yuki98}
indicating that at low energies the penetration probability for
loosely bound systems, such as the deuteron, can noticeably exceed
the conventional estimates.

The problems of tunneling and reflection of a composite particle
were discussed recently with the help of various models
\cite{saito95,sato02,kimura02,goodvin05a,goodvin05b,FZ05,bacca06}.
It was stressed that new, usually ignored, effects are important for
nuclear fusion and fission, nucleosynthesis in stars, molecular
processes, transport phenomena in semiconductors and
superconductors, both in quasi-one-dimensional and three-dimensional
systems. The resonant tunneling associated with the intrinsic
excitation, finite size effects, polarizability of tunneling
objects, evanescent modes near the barrier, real
\cite{takigawa99,BPZ99} and virtual \cite{FZ99} radiation processes
are the examples of interesting new physics. Below we consider
simple models which illustrate how ``hidden" degrees of freedom can
show up in the process of tunneling leading to a considerable
enhancement of the probability of this process.

Let the tunneling particle possess two degenerate intrinsic states
and the incident wave comes to the barrier in a pure state ``up" (it
is convenient to use the spin-1/2 language with respect to the
$z$-representation). We assume one-dimensional motion with the
simplest rectangular potential barrier of height $U_{0}$ located at
$0<x<a$. At low energy $E\ll U_{0}$, when the imaginary action
$\kappa_{0}a=[2m(U_{0}-E)]^{1/2}a$ is very large, the transmission
coefficient $T_{0}\propto \exp(-2\kappa_{0}a)$ is exponentially
small. This probability can be exponentially enhanced by a weak
``magnetic" field applied in the area of the barrier. We assume that
the interaction of this field with the particle is $-h\sigma_{x}$,
where $h$ is proportional to the transverse magnetic field.

Indeed, this field creates the ``down" spin component and splits the
states inside the barrier according to the value of $\sigma_{x}$. In
the $z$-representation the regions $x\leq 0$ and $x\geq a$ acquire
the down component in the reflected and transmitted waves,
\begin{equation}
\psi_{1}(x)=\left(\begin{array}{c}
           Ae^{ikx}+Be^{-ikx}\\
           B'e^{-ikx} \end{array}\right), \quad x\leq 0,   \label{1}
\end{equation}
and
\begin{equation}
\psi_{3}(x)=\left(\begin{array}{c}
 F\\
 F'\end{array}\right)e^{ik(x-a)}, \quad x\geq a.            \label{2}
\end{equation}
Inside the barrier the two tunneling components have slightly
different imaginary momenta, $\kappa=[2m(U_{0}-E)-h]^{1/2}$ and
$\kappa'=[2m(U_{0}-E)+h]^{1/2}$. Correspondingly, the spinor wave
function under the barrier is given by
\begin{equation}
\psi_{2}(x)=\left(\begin{array}{c}
 Ce^{-\kappa x}+C'e^{-\kappa'x}+De^{\kappa x}+D'e^{\kappa'x}\\
 Ce^{-\kappa x}-C'e^{-\kappa'x}+De^{\kappa x}-D'e^{\kappa'x}
\end{array}\right), \quad 0<x<a.                    \label{3}
\end{equation}

Performing the matching of the wave function, we find the
transmission coefficients: for spin up
\begin{equation}
T_{+}=4\frac{|\Phi+\Phi'|^{2}}{|\Phi|^{2}|\Phi'|^{2}}, \label{4}
\end{equation}
and for spin down (initially not present)
\begin{equation}
T_{-}=4\frac{|\Phi-\Phi'|^{2}}{|\Phi|^{2}|\Phi'|^{2}}, \label{5}
\end{equation}
where
\begin{equation}
\Phi(\kappa)=\left(1-\frac{ik}{\kappa}\right)\left(1-\frac{\kappa}{ik}\right)
e^{\kappa a}+\left(1+\frac{ik}{\kappa}\right)\left
(1+\frac{\kappa}{ik}\right) e^{-\kappa a},       \label{6}
\end{equation}
and $\Phi'=\Phi(\kappa')$. Assuming small penetrability, $\kappa
a\gg 1, \;\kappa' a\gg 1$, and ignoring exponentially small terms,
we obtain
\begin{equation}
T_{\pm}=4k^{2}\frac{|\kappa'(\kappa-ik)^{2}e^{-\kappa' a}\pm\kappa
(\kappa'-ik)^{2}e^{-\kappa a}|^{2}}{(\kappa^{2}+k^{2})^{2}
(\kappa'^{2}+k^{2})^{2}}.                        \label{7}
\end{equation}
If, in addition, the splitting is large enough and
$e^{-\kappa'a}\ll e^{-\kappa a}$,
\begin{equation}
T_{+}\approx
T_{-}=\frac{4k^{2}\kappa^{2}}{(\kappa^{2}+k^{2})^{2}}\,
e^{-2\kappa a}.                                \label{8}
\end{equation}
This should be compared with the transmission without splitting,
\begin{equation}
T_{0}=\frac{16k^{2}\kappa_{0}^{2}}{(\kappa_{0}^{2}+k^{2})^{2}}\,
e^{-2\kappa_{0} a}.                          \label{9}
\end{equation}
This is valid if the splitting $\Delta=\kappa_{0}-\kappa$ is large
enough in the exponents, $2a\Delta \gg 1$,
\begin{equation}
\frac{T_{\pm}}{T_{0}}\approx \frac{1}{4}\,e^{2a\Delta}, \label{10}
\end{equation}
that allows one to neglect the exponents $e^{-\kappa' a}$. The
validity condition, therefore, is that not only $\kappa_{0}a\gg 1$
but, much more strongly, $a\Delta\gg 1$, which is hard to satisfy
with electrons and realistic magnetic fields in the laboratory:
\begin{equation}
\frac{|h|}{2(U_{0}-E)}\gg \frac{1}{\kappa_{0}a} \label{11}
\end{equation}
for an electron with $a=10$ nm, $U_{0}-E=1$ eV, requires
$\kappa_{0}a\approx 50$, so that it should be $|h|/(U_{0}-E)> 1/25$,
or $|h|>0.04$ eV. However, a similar situation can be realized in
the case of a system with the ground state as a combination of two
configurations slightly split by their coupling; this splitting
plays the role of the magnetic field.

\begin{figure}[ptb]
\begin{center}
\includegraphics[
height=2.2in, width=5.5in ] {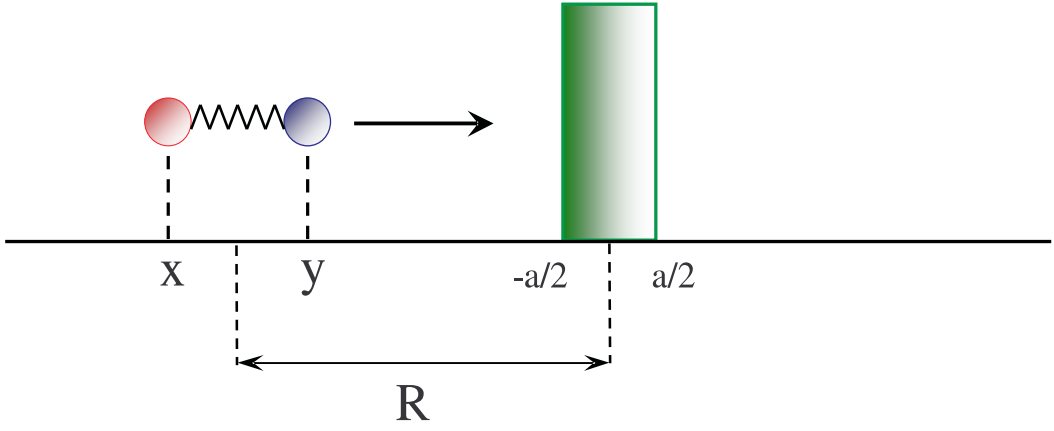}
\end{center}
\caption {Quantum tunneling of a composite particle
through a barrier. The barrier might be visible by only one
of the particles, e.g. the neutron within a deuteron does not
see a Coulomb barrier due to presence of another nucleus.}
\end{figure}

There is no need for an external magnetic field if there exists
another intrinsic state of the tunneling object that would be able
to tunnel with a larger probability. In distinction to the cases of
resonant tunneling discussed in the literature the situation is
possible when the composite particle energetically cannot be
transferred to the state with favorable conditions for tunneling.
However, even the virtual admixture of such an intermediate state
can increase penetrability. In such a case outside of the barrier
the trace of the evanescent state can exist only in the vicinity of
the barrier. Similar virtual states emerge with necessity in
reflection of a composite particle \cite{sato02} as well as in the
situation when only one of the constituents interacts with the
barrier while the rest of the constituents do not feel it
\cite{zakhariev64,matthews99,ahsan07}. This happens for example at
the Coulomb barrier for a system that contains neutral and charged
constituents (see figure 1).

Next we briefly consider the case when the excited intrinsic state
with energy $E'>E$ can virtually transfer excitation into
translational energy (the system then is still under the barrier).
The two-component wave function with the low component describing
the new intrinsic state can be written in a form similar to eqs.
(\ref{1}-\ref{3}), with simple substitutions
\begin{equation}
B'e^{-ikx}\Rightarrow B'e^{\lambda x}, \quad F'e^{ik(x-a)}
\Rightarrow F'e^{-\lambda(x-a)},                  \label{12}
\end{equation}
where $\lambda=[2m(E'-E)]^{1/2}$ gives the decay of the virtual
wave function in free space. The wave function inside the barrier
still keeps the form (\ref{3}) with $\kappa'$ now describing the
imaginary momentum of the virtual state.

The transmission coefficient here is found as
\begin{equation}
T=16k^{2}\,\left[\frac{\kappa(\kappa'+\lambda)^{2}e^{-\kappa a}+
\kappa'(\kappa+\lambda)^{2}e^{-\kappa' a}}{\left(2\kappa\kappa'+
\lambda(\kappa+\kappa')\right)^{2}+k^{2}(\kappa+\kappa'+2\lambda)^{2}}
\right]^{2},                                    \label{13}
\end{equation}
where, analogously to eq. (\ref{7}) we neglect exponentially small
terms. This expression has a simple $\lambda$-independent limit
for large excitation energy, $\lambda\gg\kappa,\kappa'$, which,
under the assumption $\exp(\kappa'a)\ll\exp(\kappa a)$, gives
\begin{equation}
T\approx \left[\frac{4k\kappa'}{(\kappa+\kappa')^{2}+4k^{2}}
\right]^{2}\,e^{-2\kappa'a}.                     \label{14}
\end{equation}
This limiting result corresponds to the sudden breakup of the
incident wave function by the edge of the barrier that is equivalent
to its instantaneous expansion into two components one of which has
a strong enhancement of the tunneling probability.

In a more realistic description, the intrinsic wave function of the
slow composite particle will change smoothly along its path. We
consider a one-dimensional motion of the complex of two particles at
positions $(x,y)$ with masses $m_{x}$ and $m_{y}$ coupled by their
interaction $V(x-y)$ and slowly moving in an external potential,
$U$,
\begin{equation}
H=\frac{p_{x}^{2}}{2m_{x}}+\frac{p_{y}^{2}}{2m_{y}}+V(x-y)+U_{x}(x)
+U_{y}(y),                                      \label{15}
\end{equation}
where we allow the external potential $U$ to act differently on
the constituents. Introducing the center-of-mass coordinate $R$,
relative coordinate $r$, and corresponding masses $M$ and $\mu$
(reduced mass), we come to the stationary Schr\"{o}dinger equation
\begin{equation}
\left\{-\frac{\hbar^{2}}{2M}\,\frac{\partial^{2}}{\partial R^{2}}
-\frac{\hbar^{2}}{2\mu}\,\frac{\partial^{2}}{\partial r^{2}}+V(r)
+U_{x}\left(R+\frac{rm_{y}}{M}\right)
+U_{y}\left(R-\frac{rm_{y}}{M}\right)-E\right\}\Psi(R,r)=0.
                                              \label{16}
\end{equation}

At low energy we can use an adiabatic ansatz
\begin{equation}
\Psi(R,r)=\psi(R)\phi(r;R),                  \label{17}
\end{equation}
where the internal function $\phi(r;R)$ describing the smooth
evolution of relative motion parametrically depends on the slow
global variable $R$ and satisfies the instantaneous equation
\begin{equation}
\left\{-\frac{\hbar^{2}}{2\mu}\,\frac{\partial^{2}}{\partial
r^{2}}+V(r) +U_{x}\left(R+\frac{rm_{y}}{M}\right)
+U_{y}\left(R-\frac{rm_{y}}{M}\right)-\epsilon(R)\right\}
\phi(r;R)=0.                                \label{18}
\end{equation}
A loosely bound state is created by the potential $V(r)$ far away
from the barrier. As the motion in the direction of the barrier
proceeds and the wings of the relative wave function propagate in
the region of the potentials $U$, this wave function is gradually
evolving along with its energy eigenvalue $\epsilon(R)$. It was
pointed out in \cite{FZ05} that the polarizability of the tunneling
system by the field of the target may increase the penetration
probability.

\begin{figure}[ptb]
\begin{center}
\includegraphics[
height=3.4in, width=3.5in ] {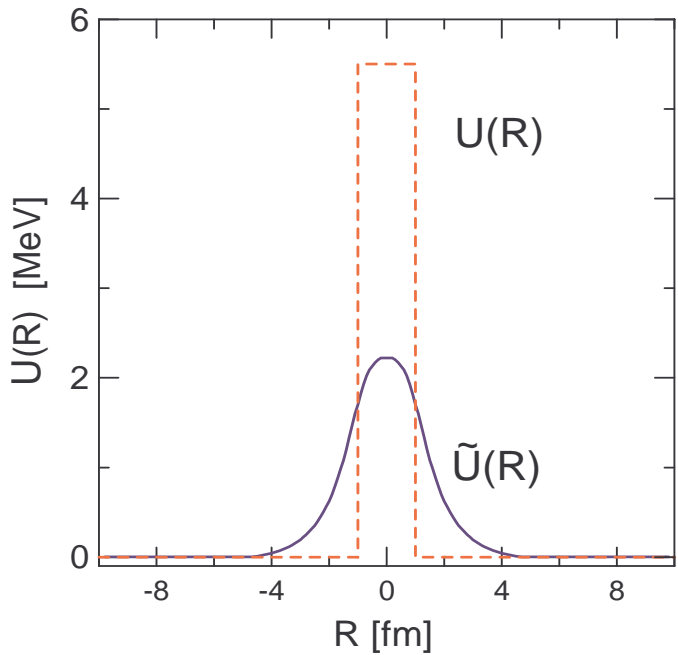}
\end{center}
\caption { The effective potential $\widetilde{U}(R)$ for the
deuteron, as obtained from eq. \ref{23}. The dashed line represents
the rectangular
potential barrier.}
\end{figure}

Multiplying eq. (\ref{16}) by $\phi^{\ast}(r;R)$, integrating out
the intrinsic variable $r$ and introducing new functions
\begin{equation}
\langle\phi|\phi\rangle=N(R), \quad
\langle\phi|\partial\phi/\partial R\rangle=N(R)\alpha(R), \quad
\langle\phi|\partial^{2}\phi/\partial R^{2}\rangle=N(R)\beta(R),
                                             \label{19}
\end{equation}
the resulting equation for $\psi(R)$ becomes
\begin{equation}
\psi^{''}+2\alpha(R)\psi'+\beta(R)\psi+
\frac{2M}{\hbar^{2}}[E-\epsilon(R)]\psi=0,       \label{20}
\end{equation}
where primes label the derivatives with respect to $R$. Defining
the new function $u(R)$,
\begin{equation}
\psi(R)=u(R)e^{-\int \alpha(R)\,dR},          \label{21}
\end{equation}
we reduce the problem to the standard Schr\"{o}dinger equation,
\begin{equation}
u''+\frac{2M}{\hbar^{2}}[E-\tilde{U}(R)]u=0     \label{22}
\end{equation}
with the effective potential
\begin{equation}
\tilde{U}{R}=\epsilon(R)-E_{0}+\frac{\hbar^{2}}{2M}\,
[\alpha^{2}(R)+\alpha'(R) -\beta(R)].        \label{23}
\end{equation}
Here the energy scale is chosen in such a way that far from the
barrier, $R\rightarrow\infty$, energy $\epsilon(R)$ coincide with
the intrinsic binding energy $E_{0}$.  Note that the solution of eq.
\ref{18} is not normalized. But eq. \ref{17} is, i.e. $\sqrt{N(R)}
\psi(R)$ is normalized to one.

The closeness of the continuum level would be dangerous in the form
of adiabatic perturbation theory where small denominators can arise.
But in the form of a differential  equation, as formulated above, we
do not throw away non-adiabatic effects. They are important and they
make the wave function to evolve. Of course, real dissociation is
impossible at low energy because  of energy conservation. When the
binding energy is small, the particles are still correlated and can
get together after the barrier. (Even in the continuum their wave
function would not be the  product of two independent plane waves,
they are still correlated because of the interaction between them.)
Our model accounts for these features.

As an application of this approach, we consider the transmission of
a composite particle through a rectangular barrier, Fig. 1, a
problem discussed for a molecule in the context of condensed matter
physics in \cite{saito94,sato02}. For nuclear applications we assume
the ``deuteron" model, when the rectangular potential of Fig. 1 acts
on one constituent (``proton") only, $U_{x}=U_{p}$, while its
partner, a ``neutron", is not influenced by the barrier, $U_{y}=0$.
The intrinsic potential $V(r)$ is also taken as a rectangular well
with parameters of depth $V_{0}=11.4$ MeV and width $r_{0}=2$ fm
reproducing the deuteron binding energy $E_{0}=-2.225$ MeV. The
choice of the barrier parameters, $U_{0}=5.5$ MeV and $a=r_{0}=2$
fm, implies that at $R=0$ the intrinsic well exactly coincides with
the barrier and the deuteron is practically unbound. The masses used
to produce the results in fig. 2 are $M=2m_N$ and $\mu=m_N/2$, where
$m_N=939$ MeV is the free nucleon mass.

The numerically calculated effective potential (23) is shown in
Fig. 2 by a solid line. It is obvious that the barrier
transmission problem is very different from that for penetration
through the original potential $U(R)$, a dashed line. At $R=0$ the
deuteron is barely bound with binding energy $\epsilon(0)=0.01$
MeV; the non-adiabatic terms [the brackets in eq. (\ref{23})] are
very small at $R=0$, so that the height of the effective potential
here is only $\tilde{U}(0)\approx 2.21$ MeV, as seen in Fig. 2.
For a deeper intrinsic potential $V(r)$, the height $\tilde{U}(0)$
would be closer to the top of the barrier but the smearing effect
of weak deuteron binding is significant.

\begin{figure}[ptb]
\begin{center}
\includegraphics[
height=3.5in, width=3.6in ] {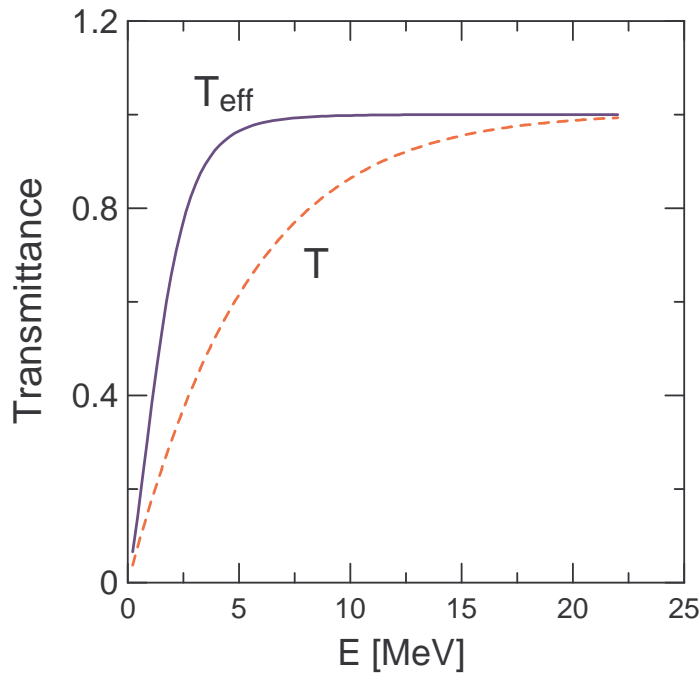}
\end{center}
\caption {Transmission probability for a deuteron incident on a
rectangular barrier (dashed-line) and on an effective potential
barrier (solid line).
Potential and energy parameters are given in the text.}
\end{figure}

The Schr\"{o}dinger equation (\ref{22}) with the symmetric effective
potential $\tilde{U}(R)$ allows, at given energy $E$, for the
solutions with definite parity, $u_{\pm}$. The solution of the
transmission problem given by the incident wave from the left is
their linear combination. We find $u_{+}$ and $u_{-}$ by a numerical
procedure starting from the center of the barrier, $R=0$, with
\begin{equation}
u_{+}(0)=1, \quad u_{+}'(0)=1, \quad u_{-}(0)=0, \quad
u_{-}'(0)=1,                               \label{24}
\end{equation}
and using an arbitrary normalization of these basic solutions.
Then we can compute the dimensionless logarithmic derivatives at a
remote point $R=R_{0}$, where $\tilde{U}(R_{0})$ is negligible,
\begin{equation}
\Lambda_{\pm}=R_{0}\,\frac{u_{\pm}'(R_{0})}{u_{\pm}(R_{0})}.
                                              \label{25}
\end{equation}
The logarithmic derivatives at the mirror point $R=-R_{0}$ are
$-\Lambda_{\pm}$.

Now we can perform the matching at $R=\pm R_{0}$ for the
scattering (transmission) problem with energy
$E=\hbar^{2}k^{2}/2M$, where the wave function is
\begin{equation}
u(R)=\left\{\begin{array}{cc} e^{ikR}+Be^{-ikR}, & -\infty<R\leq
-R_{0},\\
C_{+}u_{+}(R)+C_{-}u_{-}(R), & -R_{0}\leq R\leq R_{0},\\
Fe^{ikR}, & R_{0}\leq R<\infty.\end{array}\right.  \label{26}
\end{equation}
This gives the reflection and transmission amplitudes,
\begin{equation}
B=-\frac{1}{2}\left[\frac{\Lambda_{+}+iq}{\Lambda_{+}-iq}+
\frac{\Lambda_{-}+iq}{\Lambda_{-}-iq}\right]e^{-2iq}, \quad
F=-\frac{1}{2}\left[\frac{\Lambda_{+}+iq}{\Lambda_{+}-iq}-
\frac{\Lambda_{-}+iq}{\Lambda_{-}-iq}\right]e^{-2iq},
                                                 \label{27}
\end{equation}
where $q=kR_{0}$. The transmission probability is given by
\begin{equation}
T=\frac{q^{2}(\Lambda_{+}-\Lambda_{-})^{2}}{(q^{2}+\Lambda_{+}^{2})
(q^{2}+\Lambda_{-}^{2})}.                       \label{28}
\end{equation}
Such a calculation contains a continuous transition to global energy
exceeding the height of the effective barrier. However, then one
need to take into account the opening of the breakup channels.

The results are shown in Fig. 3 in comparison with the simple
tunneling calculation (\ref{9}) that did not take into account the
gradual adjustment of the internal wave function. A considerable
enhancement of the tunneling probability is evident.

Of course this is just a simple model that does not pretend to give
a realistic quantitative description of tunneling for a composite
object. However, we believe that it is worthwhile to point out that
there exist quantum-mechanical effects which are not discussed in
textbooks and which could be seen even in such a simplified example.
Moreover, this simple description allows us to clearly demonstrate
physics of the process not overshadowed by cumbersome computations.

We use a trial wave function that corresponds to slow motion of the
object as a whole at very low energy when dissociation channels are
forbidden. In the framework of this variational approach we solve
the problem exactly taking into account non-adiabatic corrections
(derivatives of the function describing slow motion) which are
usually neglected in a standard Born-Oppenheimer approximation in
molecular or solid state physics. This leads to the differential
Schroedinger-type equation that is solved exactly (see also
\cite{aur98}).

In conclusion, we would like to stress that tunneling of composite
objects is an important topic, regrettably not studied in detail.
Numerous applications to nuclear, atomic, molecular and condensed
matter physics,as well as to astrophysical reactions, make the
progress in understanding this problem absolutely necessary.

\begin{acknowledgments}
This work was partially supported by the NSF grant PHY-0555366
and the U.S. Department of
Energy under contract No.
DE-AC05-00OR22725,  and DE-FC02-07ER41457 (UNEDF, SciDAC-2).
\end{acknowledgments}


\begin{thebibliography}{9}

\bibitem{balantekin98}A.B. Balantekin and N. Takigawa, Rev. Mod.
Phys. {\bf 70}, 77 (1998).

\bibitem{saito94} N. Saito and Y. Kayanuma, J. Phys.: Condens.
Matter {\bf 6}, 3759 (1994).

\bibitem{arzhan91} A.V. Arzhannikov, G.Ya. Kezerashvili, Phys. Lett.
{\bf A156}, 514 (1991).

\bibitem{yuki98} H. Yuki, J. Kasagi, A.G. Lipson, T. Ohtsuki,
T. Baba, T. Noda, B.F. Lyakhov, and N. Asami, Pis'ma Zh. Eksp.
Teor. Fiz. {\bf 68}, 785 (1998) [JETP Lett. {\bf 68}, 823 (1998)].

\bibitem{saito95} N. Saito and Y. Kayanuma, Phys. Rev. B {\bf 51},
5453 (1995).

\bibitem{sato02} T. Sato and Y. Kayanuma, Europhys. Lett. {\bf
60}, 331 (2002).

\bibitem{kimura02} S. Kimura and N. Takigawa, Phys. Rev. C {\bf 66},
024603 (2002).

\bibitem{goodvin05a} G.L. Goodvin and M.R. Shegelski, Phys. Rev. A
{\bf 71}, 032719 (2005).

\bibitem{goodvin05b} G.L. Goodvin and M.R. Shegelski, Phys. Rev. A
{\bf 72}, 042713 (2005).

\bibitem{FZ05} V.V. Flambaum and V.G. Zelevinsky, J. Phys. G: Nucl.
Part. Phys. 31, 355 (2005).

\bibitem{bacca06} S. Bacca and H. Feldmeier, Phys. Rev. C {\bf 73},
054608 (2006).

\bibitem{takigawa99} N. Takigawa, Y. Nozawa, K. Hagino, A. Ono,
and D. M. Brink, Phys. Rev. C {\bf 59}, R593 (1999).

\bibitem{BPZ99} C.A. Bertulani, D.T. de Paula, and V.G.
Zelevinsky, Phys. Rev. C {\bf 60}, 031602 (1999).

\bibitem{FZ99} V.V. Flambaum and V.G. Zelevinsky,  Phys. Rev.
Lett. {\bf 83}, 3108 (1999).

\bibitem{zakhariev64} B.N. Zakhariev and S.N. Sokolov, Ann. d.
Phys. {\bf 14}, 229 (1964).

\bibitem{matthews99} M. Matthews, A. Sakharuk and V. Zelevinsky,
BAPS {\bf 44}, No. 5, p. 37 (1999).

\bibitem{ahsan07} N. Ahsan and A. Volya, BAPS {\bf 52}, No. 3, p.
159 (2007).

\bibitem{aur98} A. Bulgac, G. Do Dang and D. Kusnezov, Phys.
Rev. {\bf E 58}, 196 (1998).

\end{thebibliography}
\end{document}